# Cluster Density Matrix Embedding Theory for Quantum Spin Systems

Zhuo Fan[1, 2] and Quan-lin Jie[*, 1]

[1]Department of Physics, Wuhan University, Wuhan 430072, People's Republic of China

[2]School of Biomedical Engineering, Hubei University of Science and Technology, Xianning 437100, People's Republic of China

**Abstract**

We applied cluster density matrix embedding theory, with some modifications, to a spin lattice system. The reduced density matrix of the impurity cluster is embedded in the bath states, which are a set of block-product states. The ground state of the impurity model is formulated using a variational wave function. We tested this theory in a two-dimensional (2-D) spin-1/2 $J_1 - J_2$ model for a square lattice. The ground-state energy (GSE) and the location of the phase boundaries agree well with the most accurate previous results obtained using the quantum Monte Carlo and coupled cluster methods. Moreover, this cluster density matrix embedding theory is cost-effective and convenient for calculating the von Neumann entropy, which is related to the quantum phase transition.

---

[*] corresponding author: qljie@whu.edu.cn

# I. INTRODUCTION

The central difficulty of solving a many-body system is that the number of degrees of freedom exponentially increases with the size of the system. Most quantum many-body systems cannot be solved exactly, with the exception of a few simple cases. Many numerical simulation approaches have been proposed to resolve this difficulty in various ways. However, in many cases, it is not necessary to address all degrees of freedom. For many lattice systems with spatial symmetries, the degrees of freedom of the subsystem can capture the most important properties of the entire system. For example, for a system with long-range order, the properties of the basic cell can be used to infer the properties of the system, whereas for a system with long-range fluctuations, the reduced density matrix of the subsystem contains a large amount of valuable information. However, obtaining the reduce density matrix of the subsystem in a many-body problem remains a challenging task.

A typical method that reduces the number of degrees of freedom is dynamical mean-field theory (DMFT) [1-8]. This method maps a lattice system onto an impurity model with self-consistent bath states, which is represented by a matrix of impurity hybridizations. Following the self-consistency condition for DMFT, the impurity Green's function reproduces the local Green's function of the lattice through an effective mean field. In other words, this method treats the impurity degrees of freedom exactly and approximates the bath states at the mean-field level. The degrees of freedom of the subsystem (impurity), namely, the local Green's function, represent the system properties.

Very recently, density matrix embedding theory (DMET) [9-11], an alternative to DMFT, has been proposed to compute frequency-independent quantities in Hubbard models. In this method, the reduced density matrix of the impurity is embedded in a bath state that consists of a single bath site per impurity site. The exact embedding of the Hamiltonian is replaced with one that is exact for a one-particle mean-field lattice Hamiltonian. This method reproduces the ground-state energy curves with high accuracy compared with the Bethe ansatz for the one-dimensional (1-D) Hubbard model and compared with quantum Monte Carlo methods for the 2-D case. In general, the embedded theory divides the degrees of freedom of the lattice system into two components, with the impurity treated exactly and the bath treated approximately.

To date, the embedded theory has been used predominantly in fermion models. DMFT has been used successfully in density functional theory and typical band-structure calculations as well as for determining the electronic structures of strongly correlated materials. DMET yields highly accurate results for the ground-state energy (GSE) and correlation function in the Hubbard model. However, in only a few studies [12] has a spin system been mapped onto an impurity model. Furthermore, the treatment of the bath state in Ref [9] is not suitable for a spin lattice system. These difficulties motivate us to study how the embedded theory can be used in a spin lattice system.

In this study, we apply DMET, with some modifications, to a spin lattice system. We use a spin-1/2 antiferromagnetic $J_1 - J_2$ model for a square lattice as an example to demonstrate the use of DMET. It is straightforward to extend this method to other

2-D quantum spin systems on typical types of lattices (triangular, honeycomb, kagomé). The antiferromagnetic $J_1 - J_2$ model is the canonical model for studying the interplay of frustration effects and quantum phase transitions, and this model has attracted considerable attention over the past two decades [13-33]. It is well accepted that the ground state of this model exhibits two long-range-ordered phases, namely, a Néel phase for $J_2 < 0.4$ and a collinear phase for $J_2 > 0.6$, separated by a disordered quantum paramagnetic (QP) phase. The primary interest in the antiferromagnetic $J_1 - J_2$ model is focused on the properties of the intermediate phase corresponding to $0.4 < J_2 < 0.6$, which remain a topic of debate. The results of series expansions[14], large-N expansions[15], the projected entangled-pair states method[29], and the coupled cluster method[28] are believed to indicate the emergence of a valence bond state. By contrast, the density matrix renormalization group[32] and spin-wave calculations[13] suggest that the intermediate phase is a spin liquid state. The Hamiltonian of this model is given by

$$H = J_1 \sum_{\langle i,j \rangle} \mathbf{S}_i \cdot \mathbf{S}_j + J_2 \sum_{\langle i,k \rangle} \mathbf{S}_i \cdot \mathbf{S}_k , \qquad (1)$$

where $J_1$ and $J_2$ are the (positive) nearest-neighbor (NN) coupling and next-NN (NNN) coupling, respectively. The sums $\langle i, j \rangle$ and $\langle i, k \rangle$ run over the NN and NNN pairs, respectively. We set $J_1 \equiv 1$ in the following calculation. The GSE is determined by the bond energy $\langle \Psi | \mathbf{S}_i \cdot \mathbf{S}_j | \Psi \rangle$; therefore, we use the cluster embedding scheme, which is similar to cluster-DMFT.

The remainder of this study is organized as follows. In Sec. II, we demonstrate the application of cluster-DMET to the 2-D antiferromagnetic $J_1 - J_2$ model for a square lattice and demonstrate the optimization of the impurity wave function. In Sec. III, we present the GSE for $J_2=0$ and GSE curves for $J_2 \neq 0$. These results agree well with those obtained using several other methods. Moreover, we demonstrate the embedding effect of the approximate bath states. The reduced density matrix of the impurity spin cluster can be easily calculated using this cluster-DMET approach. The von Neumann entropy is used to determine the quantum phase transitions. Finally, a summary is presented in Sec. IV.

## II. CLUSTER DENSITY MATRIX EMBEDDING THEORY

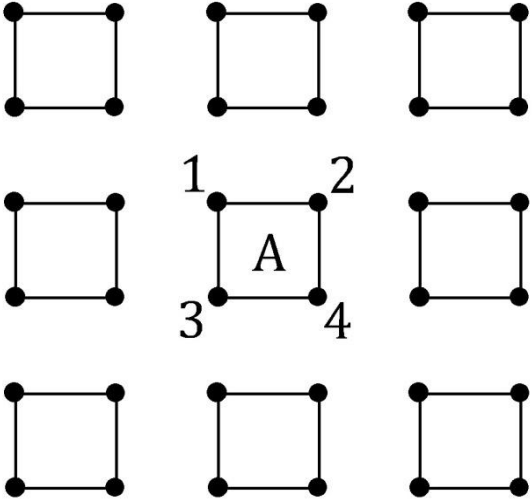

FIG. 1 A square lattice under periodic boundary conditions is divided into $2 \times 2$ spin clusters, where A is treated as an impurity cluster and the remaining spin clusters are treated as bath states.

When a spin lattice system is mapped to an impurity problem, the ground-state wave function $|\Psi\rangle$ can always be expressed as:

$$|\Psi\rangle = \sum_i a_i |\alpha\rangle_i |\beta\rangle_i ,$$ (2)

where the $\{|\alpha\rangle_i\}$ denote the basis sets of the impurity spin sites, the $\{|\beta\rangle_i\}$ denote the states of the remaining lattice sites, and the $a_i$ represent the expansion coefficients. Here, the bath states represent an exact embedding for the impurity spin cluster. However, the problem of obtaining the bath states remains a many-body problem. Following the basic concept of the embedded theory, we must use an approximate bath state to replace the exact embedding bath states. However, if we follow the original DMET as presented in Ref [9], replacing the bath states with a one-particle mean-field state, the result is a hierarchical mean field [34]. Thus, the treatment of the bath states that applies for the Hubbard model is not suitable for the spin system. We must find new alternative methods of approximating the bath states for the spin lattice system. The interaction between the impurity cluster and the bath sites should be contained in the impurity wave function. Thus, we replace the exact embedding bath state $|\beta\rangle_i$ with a set of block-product states $|\beta_{BPS}\rangle_i$. The wave function of the impurity model is

$$|\Psi\rangle_{imp} = \sum_i a_i |\alpha\rangle_i |\beta_{BPS}\rangle_i .$$ (3)

The number of block-product states equals the number of basis sets of the impurity cluster, thereby indicating that the bath states are approximated beyond the mean-field level. Then, the impurity ground-state wave function is easily computed using the linear iteration optimization algorithm [35, 36].

For simplicity, we divide a finite square lattice under periodic boundary conditions into $2 \times 2$ spin clusters, as shown in Fig. 1, and arbitrarily select one cluster A as the impurity cluster. The remaining spin clusters are treated as bath states. The shape is selected such that the equivalence of all sites in the impurity cluster is preserved. The wave vectors of the Néel state and the collinear state are $(\pi, \pi)$ and $(\pi, \pi)$ (or $(0, \pi)$), respectively. The $2 \times 2$ spin clusters match the C4 rotational symmetry of the Néel state and the C2 rotational symmetry of the collinear state. In Ref [34], $2 \times 2$ spin clusters were demonstrated to be suitable for application to a square $J_1 - J_2$ lattice. It is clear that considering a larger spin cluster could improve the results. However, the number of bath states exponentially increases with the size of the spin cluster. The ground-state wave function of the impurity model is given by

$$\left|\Psi\right\rangle_{\text{imp}} = a_0 \left|0000\right\rangle \left|\beta_{\text{BPS}}\right\rangle_0 + a_1 \left|0001\right\rangle \left|\beta_{\text{BPS}}\right\rangle_1 + \cdots + a_{15} \left|1111\right\rangle \left|\beta_{\text{BPS}}\right\rangle_{15} \qquad (4)$$

where $\left|\beta_{\text{BPS}}\right\rangle = \prod_{i=\text{bath clusters}} \left(b_{i0} \left|0000\right\rangle + b_{i1} \left|0001\right\rangle + \cdots + b_{i,15} \left|1111\right\rangle\right)$. The approximation of the bath state is identical to the hierarchical mean-field theory with spin cluster $2 \times 2$. However, we use 16 block product states to represent the bath state. We will demonstrate in the following that these block product states reproduce the interaction between the impurity sites and bath sites.

Now, the ground-state wave function is determined by a set of block-product states $\{\left|\beta_{\text{BPS}}\right\rangle_i\}$, and the expansion coefficients $\{a_i\}$ can be obtained using the linear iteration optimization algorithm. The core concept of this algorithm is to optimize the impurity wave function such that the eigenvalue of the impurity model $E_{\text{imp}} = \left\langle \Psi \left| H \right| \Psi \right\rangle_{\text{imp}}$ tends toward a minimum. For each step of iteration, we optimize one spin cluster of the block-product states, the variational parameters contain

normalization coefficients of the bath states and original parameters of the spin cluster.

We first randomly generate a wave function of the impurity model. The $k$th cluster of the $i$th block-product states of the bath states is optimized as follows. First, we express the $i$th block-product states $\alpha_i \beta_i$ as:

$$\alpha_i \beta_i = b_0 \left| 0 \; 0 \; 0 \; 0 \right\rangle \alpha_i \gamma + b_1 \left| 0 \; 0 \; 0 \; 1 \right\rangle \gamma + \cdots + b_{15} \left| 1 \; 1 \; 1 \; 1 \right\rangle_{ik}, \qquad (5)$$

where $\gamma = \prod_{j \neq k} \left( b_{ij0} \left| 0000 \right\rangle + b_{ij1} \left| 0001 \right\rangle + \cdots + b_{ij15} \left| 1111 \right\rangle \right)$. Then, the matrix elements of the Hamiltonian are calculated in the subspace spanned by $\{ \alpha_0 \beta_0, \alpha_1 \beta_1, \alpha_{i-1} \beta_{i-1}, \cdots, \alpha_{i+1} \beta_{i+1}, \alpha_{15} \beta_{15}, \left| 0000 \right\rangle \alpha_i \gamma, \left| 0001 \right\rangle \alpha_i \gamma, \cdots, \left| 1111 \right\rangle \alpha_i \gamma \}$. The eigenstate $\{ b_0, b_1, \cdots, b_{30} \}$ and the corresponding eigenvalue $E_{\text{imp}}$ can be obtained by solving the generalized eigenvalue problem. The normalization coefficients and original parameters of the $k$th cluster of the $i$th block-product states $\{ a_0, a_1, \cdots, a_{i-1}, a_{i+1}, \cdots, a_{15}, b_{ik0}, b_{ik1}, \cdots, b_{ik,15} \}$ are replaced with $\{ b_0, b_1, \cdots, b_{30} \}$. By optimizing the spin clusters of the bath states individually and repeating this iterative procedure until the eigenvalues converge, we can obtain the ground-state wave function and GSE of this impurity model. The rate of convergence depends on the truncation error $\tau = \Delta E / E$. We take $\tau = 10^{-6}$ to ensure the convergence of the ground-state wave function. The number of iterations used in our calculation is generally approximately $10^3$, which can be easily performed by a PC.

In general, our method offers several features that are different from those of the original DMET in Ref [9]. First, we use a set of block-product states to approximate the bath states. Second, we use the variational method to optimize the wave function of the impurity model. This implementation is stable and efficient.

## III. RESULTS

In the following, we discuss how to extract the bulk properties from the ground state of the impurity model. We focus primarily on the GSE. Note that the GSE of the spin lattice system is determined by the bond energy. For $J_2 = 0$, the $J_1 - J_2$ model simply reduces to the 2-D Heisenberg model, which exhibits rotational and translational invariance. All NN bond energies are identical. Therefore, we use the bond energy $e'_i = 2\langle \Psi | S_1 S_2 | \Psi \rangle$ in the impurity cluster to represent the GSE of the 2-D Heisenberg model.

Table 1 lists the GSEs obtained by calculating the impurity model for various lattice sizes. The extrapolated result is E=-0.66942, which is very close to the previous most accurate results obtained using quantum Monte Carlo (QMC) methods (E=-0.66944) [37], the coupled cluster method (CCM) (E=-0.66936) [38], and $3^{rd}$-order spin-wave theory (E=-0.66931) [39]. Note that unlike other simulation methods, which exhibit remarkable scaling effects, the bond energy of the impurity cluster is insensitive to the lattice size. This difference implies that using cluster-DMET, we can obtain reasonable results in the thermodynamics limit by calculating only a finite lattice system. The computation cost is lower than that of other methods.

TAB. 1 GSEs of the 2-D Heisenberg model for an L×L square lattice obtained using cluster-DMET. The impurity scheme is 2×2 spin cluster. The extrapolated result,

E=-0.66942, is very close to the most accurate results obtained using other methods.

| L | 8 | 12 | 16 | ∞ | QMC | CCM |
|---|---|---|---|---|---|---|
| $E_0$ | -0.669176 | -0.669392 | -0.669412 | -0.66942 | -0.66944 | -0.66936 |

For $J_2 \neq 0$, it is not appropriate to use the inter-cluster bond energy to represent the GSE of a lattice system because the rotational and translational invariance is broken. In this case, we sum all the bond energies that link the impurity sites and take half of the resulting value:

$$e_i = \frac{1}{2} J_1 \sum_{\langle i,j \rangle} \langle \mathrm{S}_i \cdot \mathrm{S}_j \rangle + \frac{1}{2} J_2 \sum_{\langle i,k \rangle} \langle \mathrm{S}_i \cdot \mathrm{S}_k \rangle , \qquad (6)$$

where the sums $\langle i,j \rangle$ and $\langle i,k \rangle$ run over the NN and NNN pairs, respectively, of site $i$. This approach will clearly lead to a loss of precision because some bonds that link the impurity and bath states are considered. From this perspective, the ideal impurity cluster scheme employs a $4 \times 4$ spin cluster and calculates the average value of the site energies of the central $2 \times 2$ spin cluster. However, the bath states consist of $2^{16}$ block-product states in this scheme, which is difficult to optimize. In fact, the GSE calculated using (6) is $E_0 = -0.6659$ for $J_2 = 0$, which is still indicates high accuracy.

Figure 2 presents the GSE curves obtained for the $2 \times 2$ impurity cluster (empty squares) for various $J_2$. For comparison, we also present the extrapolated result obtain using the CCM (empty triangles) [28, 30]. We observe that our results agree well with those of the CCM for small $J_2$ and large $J_2$. However, the accuracy is lower in the intermediate regime. As noted above, by increasing the size of the embedded cluster, one can obtain more accurate results. Therefore, we embed a $2 \times 4$ spin cluster (empty circles) as impurity sites and calculate the average value of the site energies of the impurity cluster. The block-product states of the $2 \times 2$ spin cluster are still employed for the bath states. <u>For each step of iteration, the number of the variational parameters is 271 for this impurity scheme.</u> The accuracy improves slightly in the QP phase. Note that the $2 \times 4$ impurity cluster results degrade in the collinear phase because this cluster no longer exhibits four-fold lattice symmetry. Moreover, the equivalence of the sites in the $2 \times 4$ impurity cluster is broken. This shape of the impurity cluster causes the accuracies of the impurity sites to be unequal from each other. We can calculate the average energy of the central $2 \times 2$ spin sites, representing half of the total cluster (full circles). For this calculation, the accuracy in the intermediate regime increases remarkably. From many previous studies [28-29, 31-32], we know that the intermediate phase exhibits no magnetic order but does exhibit long-range fluctuations, unlike the Néel and collinear phases. This finding may be the reason that the accuracy is lower in the QP phase for the $2 \times 2$ impurity cluster.

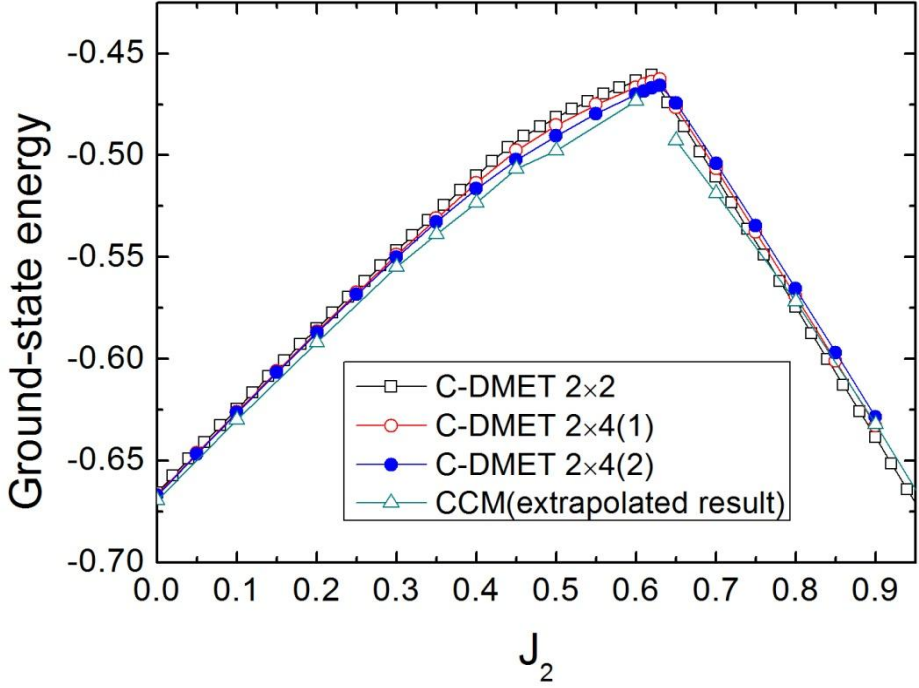

FIG. 2 (Color online) Energy curves for various values of $J_2$; the impurity schemes are a 2 ×2 spin cluster (empty squares), a 2 ×4 spin cluster (empty circles), and only the central (2 ×2) half of the 2 ×4 spin cluster (full circles); the reference data are extrapolated results obtained using the CCM (empty triangles).

To illustrate the cluster-DMET approach in the spin lattice system, we present the site energies of the impurity model for $J_2 = 0$ (see Fig. 3(a)) and $J_2 = 0.5$ (see Fig. 3(b)). According to our calculations, the system is rotationally invariant in the Néel phase and the QP phase; therefore, we select only one row of the square lattice that crosses the impurity cluster. It is apparent that the site energy curves exhibit the same features for $J_2 = 0$ and $J_2 = 0.5$. Specifically, in cluster-DMET, the square lattice system is divided into three regions based on the site energy values: the impurity cluster, the adjoining region that surrounds the impurity cluster, and the remaining outer spin clusters. The impurity cluster has the lowest site energy values, which correspond to the GSE of the spin lattice system in the thermodynamic limit. The site energy of the outer spin clusters is approximately -0.58405, which is very close to the GSE obtained using the hierarchical mean-field method with a 2 × 2 cluster [34]. These results clearly demonstrate the embedding effect of the impurity model. The outer spin clusters produce a mean-field environment, and the adjoining region mimics an exact embedding of the impurity cluster. It is known that DMFT maps a lattice system onto an impurity model. The Hamiltonian of the impurity model is divided into three components: the non-correlated bath sites, the impurity sites, and the hybridizations between the impurity and bath states. In the original DMET, the impurity Hamiltonian is approximated as:

$$H = H_A + h_{AB} + h_B,$$ (7)

where $H_A$ represents the exact embedded impurity and $h_B$ represents the mean-field embedding of the bath Hamiltonian. The form of this Hamiltonian suggests that the site energy obtained using our variational wave function is identical to the approximations of DMFT and the original DMET. The accuracy of the impurity site energy is dependent on the effective width of the square ring surrounding the impurity sites. Because of the mean-field approximation of the bath state, the impurity site energy is insensitive to the lattice size.

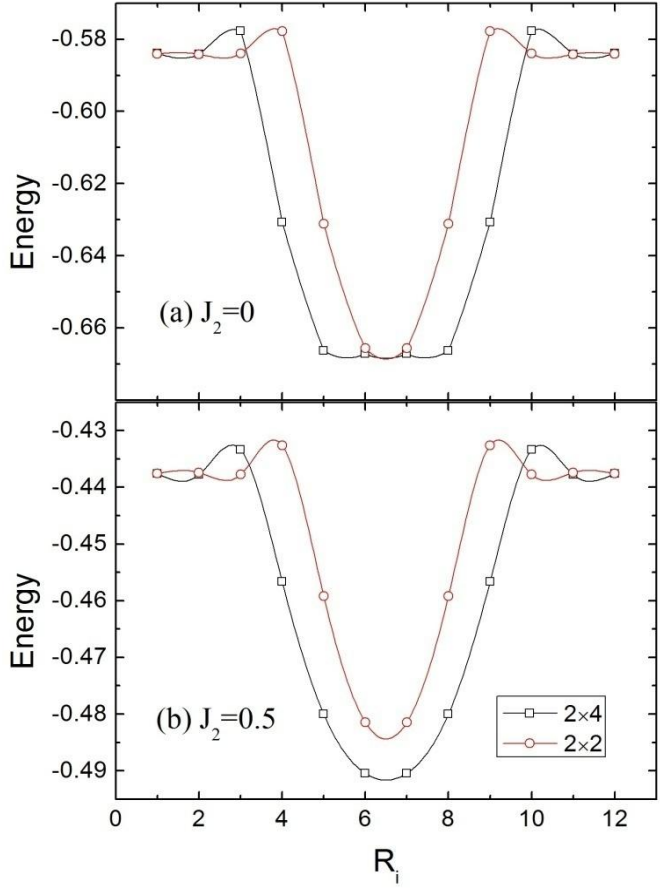

FIG. 3 (Color online) The position distributions of site energies for $J_2 = 0$ (a) and $J_2 = 0.5$ (b) obtained using cluster-DMET with $2 \times 4$ (squares) and $2 \times 2$ (circles) impurity clusters. $R_i$ is the position of site $i$; the lowest values of the curves correspond to the impurity sites; the surrounding spins produce an embedding effect on the impurity cluster; and the remaining bath spins produce a $2 \times 2$ cluster mean field.

In the results presented above, we observe that the $2 \times 2$ impurity scheme is sufficient to obtain the high accuracy GSE for the Néel state; however, the accuracy is lower for $J_2 = 0.5$, which corresponds to the QP phase. As an attempt to address this shortcoming, we consider a larger $2 \times 4$ impurity cluster. We select one row across the long edge of the $2 \times 4$ impurity cluster. In this case, the GSEs of the four impurity sites are observed to be nearly equal for $J_2 = 0$ (see Fig. 3(a)). Compared with the $2 \times 2$ impurity scheme, the $2 \times 4$ impurity cluster does not yield any improvement to the GSE for the Néel state. However, for $J_2 = 0.5$, the GSEs of the central two sites are significantly improved in the disordered phase, whereas those of the two sites at the cluster edge are not improved (see Fig. 3(b)). In other words, although a $2 \times 4$ spin cluster is embedded in the bath states, we still treat only the central $2 \times 2$ spin cluster as an impurity. In this embedding implementation, the ground-state wave function of the cluster impurity model has the following form:

$$|\Psi\rangle = \sum_m \left( a_m |\alpha_m\rangle \sum_n u_n |\mu_n\rangle |\beta_{\text{BPS}}\rangle_{mn} \right), \tag{8}$$

where the $\{\alpha_m\}$ denote the basis set of the central $2 \times 2$ impurity cluster and the $\{\mu_n\}$ denote the basis set of the remaining four spin sites of the $2 \times 4$ spin cluster. In fact, if we repeat this procedure, we ultimately obtain the exact wave function of the lattice system. This finding can be explained by the fact that the bath states are treated more accurate, robust and accommodative. Specifically, several spin sites that surround the impurity cluster are treated exactly. However, the outer spin clusters are still block-product mean-field approximations. As observed in Fig. 3, the effect of this embedding method is to enlarge the adjoining region that mimics exact embedding. This finding suggests that the accuracy of the impurity sites is dependent on the effective width of the adjoining region surrounding the impurity sites.

We know that the Néel phase is an ordered crystal state and that the spin-spin fluctuation decays rapidly with distance; therefore, it is sufficient for the adjoining region that produces the embedding effect to consist of two rows of spins. However, the QP phase exhibits long-range fluctuations, and therefore, in this case, the adjoining region needs more spins. When we treat the central $2 \times 2$ spin sites as an impurity cluster, the remaining four spin sites at the edge of the $2 \times 4$ cluster, which will also be treated exactly, are counted as bath sites. Thus, the accuracy of the impurity site energy increases.

In general, compared with other high accuracy methods, such as density matrix renormalization group (DMRG), QMC, and CCM, this cluster-DMET approach also provides an efficient method of obtaining accurate results for a lattice system in the thermodynamic limit. The most attractive feature of this cluster-DMET approach is that the multiple block-product bath states are sufficient to obtain high accuracy GSEs for an ordered magnetic phase. The cluster-DMET calculations for optimizing the wave function are much less expensive. For a disordered state, we can use Equation (8) to make the bath states much more accurate. It offers a good compromise between accuracy and computational cost. These capabilities make cluster-DMET more powerful. Moreover, this cluster DMET works in the thermodynamic limit. The properties of impurity cluster represent the cluster sites in an infinite lattice system, although we just calculate a finite impurity model.

Although this method provides accurate GSE results, its shortcomings are obvious. Because the bath state is treated approximately, the spin-spin correlation function between two spin sites cannot be captured when one site belongs to the impurity state and the other belongs to the bath state. However, this cluster-DMET approach nevertheless affords a reasonable reduced density matrix of the impurity cluster because the bath states mimic an exact embedding. The density matrix offers the following advantage over two-site correlation: it encodes the total amount of information shared between the two subsystems. This information is quantified by the von Neumann entanglement. If the lattice spins system is partitioned into a subsystem A and its complement B, the von Neumann entanglement entropy between subsystem A and B can be defined as:

$$S(\rho_A) = -\text{Tr}(\rho_A \ln \rho_A), \tag{9}$$

where $\rho_A = \text{tr}_B |\Psi\rangle\langle\Psi|$ is the reduced density matrix. The relationship between the quantum phase transition and the von Neumann entanglement entropy has been reported for many 1-D systems [40-45]. However, few studies have focused on the von Neumann entanglement entropy of 2-D systems because of the difficulty of obtaining the reduced density matrix. Using this method, we can obtain the reduced density matrix of the impurity cluster by tracing out the bath states, which is very convenient for calculating the von Neumann entropy entanglement. We can also calculate either one-site or two-site von Neumann entropy entanglement in the impurity cluster.

As is well known [40-45], a discontinuity or singularity in the entanglement entropy indicate a first-order quantum phase transition, and a peak in the derivative of the entropy indicates a continuous quantum phase transition. Thus, we calculate the one-site, two-site, and plaquette von Neumann entropy entanglement of the impurity cluster, which are labeled in Fig. 1. Figure 4 shows several entanglement entropy values for various $J_2$: $S(\rho_1)$ (full squares), $S(\rho_{12})$ (empty squares), $S(\rho_{13})$ (empty circles), $S(\rho_{14})$ (empty triangles), and $S(\rho_{1234})$ (full diamonds). All curves exhibit a discontinuity or singularity at $J_2 = 0.62$, which corresponds to a first-order transition from the QP phase to the collinear phase. A second-order phase transition from the Néel phase to the QP phase is captured by the derivative of the entanglement entropy. All curves peak at $J_2 \approx 0.42$. The location of the phase boundaries is consistent with the observations of previous studies [28, 30-32].

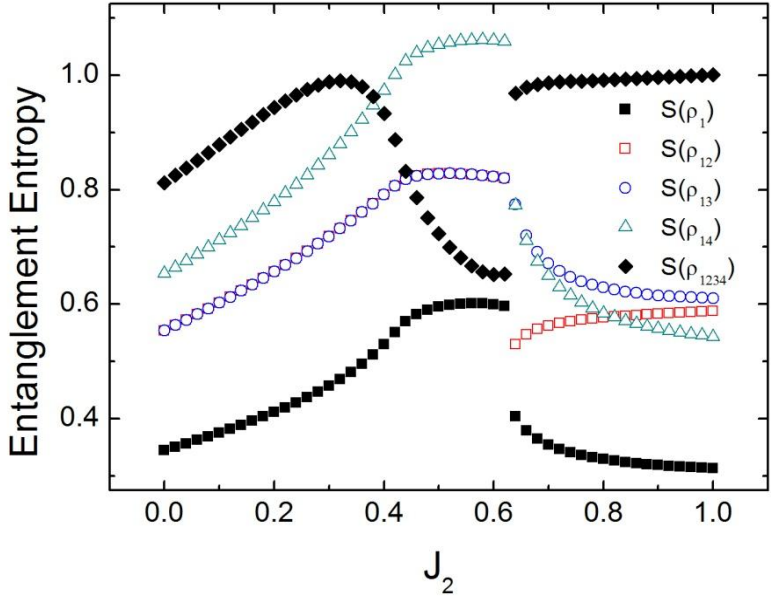

FIG. 4 (Color online) The entanglement entropy curves for various $J_2$; all curves exhibit a discontinuity at $J_2 = 0.62$, indicating a first-order transition from the QP phase to the collinear phase.

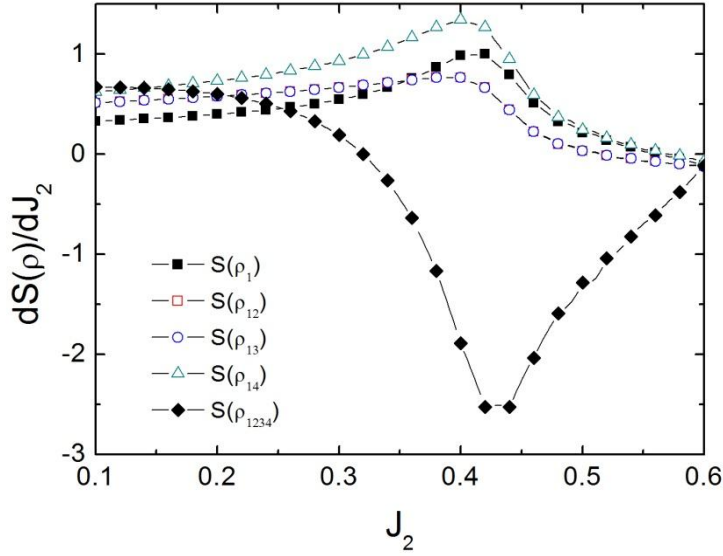

FIG. 5 (Color online) The derivative of the von Neumann entanglement entropy $dS(\rho)/dJ_2$ plotted versus $J_2$; all curves peak at $J_2 \approx 0.42$, corresponding to a second-order transition from the Néel phase to the QP phase.

Notably, the four-site block-block entanglement entropy exhibits some noticeably different features compared with the other curves presented in Fig. 4. The entanglement entropy reaches a maximum at $J_2 \approx 0.3$, which does not correspond to a phase transition. In addition, the entanglement entropy values decrease in the intermediate regime, indicating that the correlation between a plaquette and the remaining lattice sites is relatively small when $J_2$ is approximately 0.5. This result suggests that the major contribution to the entanglement of the QP phase originates from the $2 \times 2$ spin cluster. The $2 \times 2$ cluster acts as a unit in the QP phase. which may indicate that the QP phase has a weak plaquette valence solid state.

## IV. SUMMARY

In this study, we apply cluster-DMET to spin systems. In this method, a lattice system is mapped onto an impurity model. We treat the impurity spin sites exactly and the bath states as a set of block-product states that mimic an exact embedding of the impurity cluster. The bath state is optimized using a variational approach. We demonstrate this method using the 2-D spin-1/2 $J_1 - J_2$ model on a square lattice. It is straightforward to extend this cluster-DMET approach to other 2-D quantum spin systems on other types of lattices. For the considered systems, an impurity cluster with a size of $2 \times 2$ is sufficient to obtain a highly accurate GSE for the ordered Néel phase and the collinear phase, with the bath states approximated by a set of block-product states. For a disordered QP phase, we can also obtain reasonable results by enlarging the size of the impurity cluster. This cluster-DMET approach is very efficient and convenient for calculating the von Neumann entropy, which provides an entanglement-based view of the quantum phase transition.

## V. ACKNOWLEDGEMENTS

This work was supported by the National Natural Science Foundation of China (Grant No. 10875087). We acknowledge useful discussions with Cheng-bo Duan.

FIG. 1 A square lattice under periodic boundary conditions is divided into $2 \times 2$ spin clusters, where A is treated as an impurity cluster and the remaining spin clusters are treated as bath states.

TAB. 1 GSEs of the 2-D Heisenberg model for an L×L square lattice obtained using cluster-DMET. The impurity scheme is $2 \times 2$ spin cluster. The extrapolated result, E=-0.66942, is very close to the most accurate results obtained using other methods.

FIG. 2 (Color online) Energy curves for various values of $J_2$; the impurity schemes are a $2 \times 2$ spin cluster (empty squares), a $2 \times 4$ spin cluster (empty circles), and only the central (2 × 2) half of the 2 × 4 spin cluster (full circles); the reference data are extrapolated results obtained using the CCM (empty triangles).

FIG. 3 (Color online) The position distributions of site energies for $J_2 = 0$ (a) and $J_2 = 0.5$ (b) obtained using cluster-DMET with 2 × 4 (squares) and 2 × 2 (circles) impurity clusters. $R_i$ is the position of site $i$; the lowest values of the curves correspond to the impurity sites; the surrounding spins produce an embedding effect on the impurity cluster; and the remaining bath spins produce a 2 × 2 cluster mean field.

FIG. 4 (Color online) The entanglement entropy curves for various $J_2$; all curves exhibit a discontinuity at $J_2 = 0.62$, indicating a first-order transition from the QP phase to the collinear phase.

FIG. 5 (Color online) The derivative of the von Neumann entanglement entropy $dS(\rho)/dJ_2$ plotted versus $J_2$; all curves peak at $J_2 \approx 0.42$, corresponding to a second-order transition from the Néel phase to the QP phase.